\PassOptionsToPackage{unicode}{hyperref}
\PassOptionsToPackage{hyphens}{url}
\PassOptionsToPackage{dvipsnames,svgnames,x11names}{xcolor}
\documentclass[
  12pt]{article}

\usepackage{amsmath,amssymb}
\usepackage{iftex}
\ifPDFTeX
  \usepackage[T1]{fontenc}
  \usepackage[utf8]{inputenc}
  \usepackage{textcomp} 
\else 
  \usepackage{unicode-math}
  \defaultfontfeatures{Scale=MatchLowercase}
  \defaultfontfeatures[\rmfamily]{Ligatures=TeX,Scale=1}
\fi
\usepackage{lmodern}
\ifPDFTeX\else  
\fi
\IfFileExists{upquote.sty}{\usepackage{upquote}}{}
\IfFileExists{microtype.sty}{
  \usepackage[]{microtype}
  \UseMicrotypeSet[protrusion]{basicmath} 
}{}
\makeatletter
\@ifundefined{KOMAClassName}{
  \IfFileExists{parskip.sty}{%
    \usepackage{parskip}
  }{
    \setlength{\parindent}{0pt}
    \setlength{\parskip}{6pt plus 2pt minus 1pt}}
}{
  \KOMAoptions{parskip=half}}
\makeatother
\usepackage{xcolor}
\makeatletter
\ifx\paragraph\undefined\else
  \let\oldparagraph\paragraph
  \renewcommand{\paragraph}{
    \@ifstar
      \xxxParagraphStar
      \xxxParagraphNoStar
  }
  \newcommand{\xxxParagraphStar}[1]{\oldparagraph*{#1}\mbox{}}
  \newcommand{\xxxParagraphNoStar}[1]{\oldparagraph{#1}\mbox{}}
\fi
\ifx\subparagraph\undefined\else
  \let\oldsubparagraph\subparagraph
  \renewcommand{\subparagraph}{
    \@ifstar
      \xxxSubParagraphStar
      \xxxSubParagraphNoStar
  }
  \newcommand{\xxxSubParagraphStar}[1]{\oldsubparagraph*{#1}\mbox{}}
  \newcommand{\xxxSubParagraphNoStar}[1]{\oldsubparagraph{#1}\mbox{}}
\fi
\makeatother

\usepackage{longtable,booktabs,array}
\usepackage{calc} 
\usepackage{etoolbox}
\makeatletter
\patchcmd\longtable{\par}{\if@noskipsec\mbox{}\fi\par}{}{}
\makeatother
\IfFileExists{footnotehyper.sty}{\usepackage{footnotehyper}}{\usepackage{footnote}}
\makesavenoteenv{longtable}
\usepackage{graphicx}
\makeatletter
\def\maxwidth{\ifdim\Gin@nat@width>\linewidth\linewidth\else\Gin@nat@width\fi}
\def\maxheight{\ifdim\Gin@nat@height>\textheight\textheight\else\Gin@nat@height\fi}
\makeatother
\setkeys{Gin}{width=\maxwidth,height=\maxheight,keepaspectratio}
\makeatletter
\def\fps@figure{htbp}
\makeatother

\makeatletter
\@ifpackageloaded{caption}{}{\usepackage{caption}}
\AtBeginDocument{%
\ifdefined\contentsname
  \renewcommand*\contentsname{Table of contents}
\else
  \newcommand\contentsname{Table of contents}
\fi
\ifdefined\listfigurename
  \renewcommand*\listfigurename{List of Figures}
\else
  \newcommand\listfigurename{List of Figures}
\fi
\ifdefined\listtablename
  \renewcommand*\listtablename{List of Tables}
\else
  \newcommand\listtablename{List of Tables}
\fi
\ifdefined\figurename
  \renewcommand*\figurename{Figure}
\else
  \newcommand\figurename{Figure}
\fi
\ifdefined\tablename
  \renewcommand*\tablename{Table}
\else
  \newcommand\tablename{Table}
\fi
}
\@ifpackageloaded{float}{}{\usepackage{float}}
\floatstyle{ruled}
\@ifundefined{c@chapter}{\newfloat{codelisting}{h}{lop}}{\newfloat{codelisting}{h}{lop}[chapter]}
\floatname{codelisting}{Listing}

\makeatother
\makeatletter
\@ifpackageloaded{caption}{}{\usepackage{caption}}
\@ifpackageloaded{subcaption}{}{\usepackage{subcaption}}
\makeatother

\ifLuaTeX
  \usepackage{selnolig}  
\fi
\usepackage[]{natbib}
\usepackage{bookmark}

\IfFileExists{xurl.sty}{\usepackage{xurl}}{} 
\hypersetup{
  pdftitle={Title},
  pdfauthor={Author 1; Author 2},
  pdfkeywords={3 to 6 keywords, that do not appear in the title},
  colorlinks=true,
  linkcolor={blue},
  filecolor={Maroon},
  citecolor={Blue},
  urlcolor={Blue},
  pdfcreator={LaTeX via pandoc}}

\newcommand{\anon}{1}

\usepackage{multicol}
\usepackage{enumerate}

\newcommand{\bX}{\mathbf{X}}

\newcommand{\bZ}{\mathbf{Z}}

\def\hat {\widehat}

\newcommand{\bbeta}{\boldsymbol{\beta}}

\newcommand{\btheta}{\boldsymbol{\theta}}

\usepackage[ruled,vlined,lined,boxed,linesnumbered]{algorithm2e}

\usepackage{amssymb,verbatim}
\usepackage{epsfig}
\usepackage{fancyhdr}
\usepackage{pdflscape}

\usepackage{booktabs}
\usepackage{enumerate}
\usepackage{enumitem}

\usepackage{array}
\usepackage{amstext}
\usepackage{amsthm}
\usepackage{amsopn}
\usepackage{amsfonts}
\usepackage{amsmath}
\usepackage{mathrsfs}
\usepackage{amsbsy}
\usepackage{multicol}
\usepackage{colortbl}
\usepackage{multirow}
\usepackage{fancyhdr}
\usepackage{appendix}
\usepackage{pdflscape}

\theoremstyle{plain}
\newtheorem{theorem}{Theorem}[section]

\theoremstyle{definition}

\begin{document}

\def\spacingset#1{\renewcommand{\baselinestretch}%
{#1}\small\normalsize} \spacingset{1}


\if1\anon
{
  \title{\bf Semi-supervised Concordance Learning for Optimal Individual Treatment Regimes}
  \author{ Mengjiao Peng$^*$, Yong Zhou\thanks{
    The authors gratefully acknowledge \textit{ the Fundamental and Interdisciplinary Disciplines Breakthrough Plan of the Ministry of Education of China (JYB2025XDXM904), the National Natural Science Foundation of China (72331005, 12301337, 12471253), the National key research and development program (2021YFA1000101, 2021YFA1000102 and 2021YFA1000104), and Shanghai Key Program of Computational Biology (23JS1400500).}}\hspace{.2cm}\\
    Key Laboratory of Advanced Theory and Application in Statistics and Data Science,\\ MOE, School of Statistics, East China Normal University, Shanghai, China\\
    Wenbin Lu \\
    Department of Statistics, North Carolina State University, Raleigh, USA}
  \maketitle
} \fi

\if0\anon
{
  \bigskip
    \begin{center}
    {\LARGE\bf Title}
\end{center}
} \fi
\begin{abstract}
Finding the optimal individualized treatment rule that maps individual characteristics or contextual information to treatment assignments has been extensively investigated in existing literature, with widespread practical applications. This paper considers the estimation of optimal treatment regimes within a semi-supervised data framework (exemplified by electronic medical record data). In such settings, only a tiny proportion of observations have observed outcome labels, owing to high labeling costs, time limitations, data privacy concerns, and other constraints, while covariates and treatment assignments are available for all study subjects. We develop a semi-parametric inference method for optimal treatment regimes,  which leverages outcome- unlabeled samples with complete covariate and treatment information to enhance estimation efficiency. The proposed estimation framework consists of two key steps: first, flexible nonparametric imputation via single-index kernel smoothing; second, subsequent estimation of the optimal treatment regime based on concordance-assisted learning.  We establish the consistency and asymptotic normality of our proposed estimators. Numerical simulation studies demonstrate that our method achieves higher efficiency and stronger robustness relative to fully supervised estimators under finite-sample settings. We further validate the practical value of our proposed framework using the MIMIC-III and ACTG175 datasets.
\end{abstract}

\noindent%
{\it Keywords:} Optimal treatment regime; Semi-supervised inference; Nonparametric imputation
\vfill

\newpage
\spacingset{1.8} 

\section{Introduction}\label{sec-intro}
The problem of finding the optimal individualized treatment rule that maps an individual’s characteristics or contextual information to the treatment assignment has been extensively studied in the literature due to its important applications in practice, such as disease management (\cite{bertsimas2017personalized}), public policy making (\cite{kube2019allocating}), and context-based recommender system (\cite{aggarwal2016recommender}), where treatment can represent drugs, interventions, programs, strategies, policies, and so on. Especially, in precision medicine, which has received substantial scientific and commercial attention in recent years, the aim is to derive the optimal treatment regime that optimizes patients’ expected outcomes of interest (denoted by $Y$) and assigns treatment for each patient according to his/her personal information (denoted by $X$), such as their genetic, physiologic, demographic and clinical characteristics.

When investigating treatment effects, it is often the case that covariates $X$ and treatment assignments $A$ are more readily available than the outcome variable $Y$, which is typically harder to collect or more expensive to obtain. Outcome-unlabeled data, consisting of observations of $(X,A)$ without the outcome $Y$, are often abundant, while labeled data (data with covariates $X$, treatment assignments $A$, and outcome variable $Y$ observed) are relatively scarce, resulting in a complex semi-supervised (SS) data structure combining both labeled and outcome-unlabeled data \citep{chakrabortty2016robust, lan2022imputations}.
Practical data observed in biomedical research and clinical studies often exhibit a SS structure, posing significant challenges in evaluating treatment effects and inferring optimal treatment regimes \citep{gunn2024adaptive}. Especially in the era of big data, where a vast amount of outcome-unlabeled data sets, typically recorded digitally, are becoming increasingly easy to obtain and process. Like the electronic medical records (EMR) data, where obtaining validated phenotype information (outcome) can be labor-intensive and expensive, while digitally recorded data on clinical variables and treatment or exposure assignments are readily available for all subjects \citep{liao2010electronic}.

Classical methods for estimating optimal treatment regimes typically include Q-learning (Q representing ``quality"), which posits regression models for the outcome of interest (
\cite{wang2018quantile}), and A-learning  (A meaning ``advantage'') which directly builds models for the contrast functions and uses doubly robust estimating equations to estimate the contrast functions by incorporating the estimated propensity score functions (\cite{robins2000marginal}). 
Compared with Q-learning, A-learning is more robust to model misspecification.  Q- and A-learning are both indirect approaches as they rely on maximizing or minimizing an objective function to infer the optimal DTRs and thus emphasize prediction accuracy of the clinical response model instead of directly optimizing the decision rule (\cite{zhao2012estimating}). Recently, value-function-based optimization methods (\cite{zhang2012robust, zhang2013robust}; 
\cite{zhou2017residual}; \cite{mo2021learning}) were proposed for deriving the optimal treatment regime by directly maximizing the estimated mean potential outcome (the value function) under a given treatment regime. Some other machine learning techniques like direct learning (D-learning) have been also proposed to estimate optimal ITRs, see \cite{qi2018d} and \cite{qi2020multi}. In addition,  \cite{fan2017concordance} and \cite{shi2021concordance} utilized a type of concordance function for prescribing treatment and proposed the concordance-assisted learning for estimating the optimal treatment regime, based on the natural intuition that a good treatment regime shall be more likely to assign a subject to the treatment group than others, if he or she has a larger benefit of receiving treatment. The methods mentioned above for estimating the optimal treatment regimes do not take into account the outcome-unlabeled information and belong to the category of fully supervised learning methods.

In recent years, SS learning, which utilizes a mixture of a small amount of labeled data and a large amount of outcome-unlabeled data for modeling and pattern recognition, has become an exciting area of research in statistics and machine learning, with several studies investigating its potential applications (e.g. 
\cite{Angelopoulos2023science}; 
\cite{zhang2025leveraging}; \cite{zhang2025semi}; \cite{hou2025efficient}).  A typical SS setting is characterized by two types of available data: (i) a small or moderate-sized ``labeled" data set, $\mathcal{L}$, containing observations of the outcome $Y$, covariates $\mathbf{X}$, and treatment assignment $A$, and (ii) an ``outcome-unlabeled" data set, $\mathcal{U}$, of much larger size, containing observations of $(\mathbf{X},A)$ but not $Y$. Such a scenario arises naturally whenever covariates and treatment assignments are easily available, allowing for plentiful outcome-unlabeled data, but the outcome is costly or difficult to obtain, thereby limiting the size of $\mathcal{L}$. The imbalanced sample sizes of labeled and outcome-unlabeled data present significant challenges for fully supervised methods, which ignore the massive amounts of outcome-unlabeled information. The major concern of SS learning is to effectively utilize both the labeled and outcome-unlabeled data for more accurate parameter estimation (\cite{chapelle2009semi}; \cite{zhu2009introduction}). Reference works on SS learning primarily focus on classification problems (
\cite{cannings2020local}) and regression frameworks ( 
\cite{chakrabortty2018efficient}; 
\cite{Deng2020}). 

However, SS learning has not received enough attention in deriving optimal treatment regimes despite the SS structure of the rich historical information in precision medicine. Besides, the SS framework relaxes the stringent requirement in the general missing data problem, where the sample size of labeled and outcome-unlabeled data strictly needs to exceed 0. SS learning allows the proportion of fully observed observations is allowed to vanish asymptotically, resulting in significant differences from completely randomly missing data in terms of both methodological approaches and theoretical results. It is highly valuable to fully exploit and utilize the rich historical outcome-unlabeled information to improve the accuracy and efficiency of optimal treatment regime estimation due to the limited labeled data in practice. In classic SS learning, the knowledge of the distribution of the observed features $\mathcal{P}_{X,A}$ obtained from outcome-unlabeled data is combined to improve the inference of the conditional distribution $\mathcal{P}_{Y|X,A}$ (\cite{chapelle2009semi}). Unlike fully supervised learning, SS learning permits the full exploitation of rich historical outcome-unlabeled information and further enhances the estimation efficiency for optimal treatment regimes. \cite{cheng2021robust} proposed an SS estimator for the average treatment effect based on missing response imputation followed by inverse probability weighting estimation of propensity score and average treatment effect. However, their imputation method requires the knowledge of post-treatment surrogate variables that are predictive of the response, while the surrogate variables may not necessarily exist or be obtainable in medical research and clinical data.
\cite{sonabend2023semi} proposed a semi-supervised off-policy reinforcement learning framework for the optimization and evaluation of dynamic treatment regimes. Their semi-supervised learning (SSL) estimator improves efficiency by jointly utilizing labeled and outcome-unlabeled data, together with outcome surrogates, to estimate the value function. Specifically, they developed a doubly robust value function estimator based on the augmented inverse probability weighting (AIPW) approach, which remains consistent provided that either the Q-function model or the propensity score model is correctly specified.
\cite{gunn2024adaptive} utilized observed information in outcome-unlabeled data to derive the estimation of the contrast function, which in turn improves the estimation of the linear decision rule based on a semiparametric working model. 
Their method smooths the objective function directly over the covariate vector in kernel regression, which 
can become computationally intensive and suffer from the curse of dimensionality.
\cite{jiao2024smoothed} and \cite{li2025doubly} proposed various forms of semisupervised value functions based on the imputed values by single index kernel smoothing, incorporating both labeled and outcome-unlabeled data components.

When there is a large amount of outcome-unlabeled data in real-world studies, fully supervised learning methods that rely solely on a small amount of labeled data, without accounting for the information in the large amount of outcome-unlabeled data, may lead to serious bias or loss of estimation efficiency. Few SS learning methods for deriving optimal treatment regimes using outcome-unlabeled data either rely on knowledge of surrogate variables (which may not exist or may not be obtainable) or cannot handle the estimation challenges posed by high-dimensional outcome-unlabeled data. This paper aims to propose new SS learning algorithms and estimation methods that fully utilize and explore the entire dataset, including labeled and outcome-unlabeled data, to enhance the evaluation of treatment effectiveness
and improve the inference of the optimal treatment regime.
The major contribution of this paper is the proposed reliable and robust SS learning algorithm, which ensures the comprehensive utilization and exploration of all available data, thereby enhancing the overall efficiency of estimating optimal treatment regimes. 
Additionally, in situations where the propensity score function is not known, the paper presents a doubly robust SS estimation method within a specific class of monotonic index models. Our estimators are shown to be consistent and asymptotically normal using techniques including the theory of empirical process and U-statistics.

The rest of the paper is organized as follows. Section \ref{section2} introduces the data, notations, assumptions, and concordance-assisted learning (CAL). In Section \ref{section3} and \ref{section4}, we formally describe the proposed semi-supervised concordance learning with known and unknown propensity scores, respectively. Asymptotic properties of the proposed estimator are provided in Section \ref{section5}. Extensive simulation studies are conducted in  Section \ref{section6}. Results of two real data applications are provided in Section \ref{section7} and Section S4 in the supplementary material, respectively. Some concluding remarks are given in Section \ref{section9}. Additional theoretical results, additional simulation results, and all technical details are relegated to the supplementary material.

\section{Methodology} \label{section2}
\subsection{Data representation }
Denote $Y \in \mathbb{R}$ as the continuous outcome variable and assume that a larger value of $Y$ implies a better response without loss of generality.
Let $\bX \in \mathbb{R}^p$ be the predictor vector with fixed dimensionality $p$. Denote $A$ as the treatment indicator, which takes values in $\mathcal{A}=\{1,0\}$. Suppose there are two treatment options, e.g., control and experimental treatment, in a clinical trial, and let A, taking values 0 or 1 in accordance with the two options, denote the observed treatment received. Let $Y^{*}(a)$ denote the potential outcome that would be observed were a subject to receive treatment $a \in \mathcal{A}$. A treatment regime $d(\bX)$ is a deterministic function that maps
$\bX \in \mathbb{R}^p$ to $a \in \mathcal{A}$. An optimal treatment regime in class $\mathcal{D}$ is defined as
\[
    d^{\text {opt }}=\arg \max _{d \in \mathcal{D}} E\left[Y^{*}\{d(\bX)\}\right],
\]
where $\mathcal{D}$ represents a class of treatment regimes under consideration. Here we focus on a general class of linear decision rules $d(\mathbf{X})=I\left(\beta^{\prime} \bX \geqslant c\right)$ for simplicity, where $\beta$ is a $ p$-dimensional vector of parameters and $c$ is a scalar. The value function $E\left[Y^{*}\{d(\bX)\}\right]$ measures the effectiveness 
under a given treatment regime $d$.  

In reality, we observe $Y^{*}(1)$ or $Y^{*}(0)$ for one subject, but never both. To evaluate the effect of treatment and estimate the value function based on observed data, we adopt the potential or counterfactual outcome framework for causal inference (\cite{rubin1978bayesian}). First, assume that the observed result is the potential outcome corresponding to the treatment the subject actually receives (consistency), that is, $Y=Y^{*}(1) A+Y^{*}(0)(1-A)$. Second, assume that $A$ and $\left\{Y^{*}(0), Y^{*}(1)\right\}$ are independent conditional on $\bX$ (no unmeasured confounders). 
Based on these two assumptions, \cite{zhang2012robust} proposed an IPSW estimator for the value function, i.e.
\begin{eqnarray}\label{m-eq:2.1}
    \hat{V}_{a}(\beta,c)=\frac{1}{n} \sum_{i=1}^{n} \frac{Y_{i} I\left\{A_{i}=I\left(\beta^{\prime} \bX_i \geqslant c\right)\right\}}{A_{i} \pi\left(\mathbf{X}_{i}\right)+\left(1-A_{i}\right)\left\{1-\pi\left(\mathbf{X}_{i}\right)\right\}},
\end{eqnarray}
where $\pi\left(\mathbf{X}_{i}\right)=P\left(A_{i}=1 \mid \mathbf{X}_{i}\right)$ is the propensity score. To find the optimal treatment regime within the class of linear decision rules, they proposed maximizing $\hat{V}_{a}(\beta, c)$ with respect to $\beta$ and $c$ subject to the constraint $\left\|\left(c, \beta^{\prime}\right)^{\prime}\right\|=1$, where $\|\cdot \|$ represent the Euclidean norm.

In the SS setting considered in this paper, the data available consist of two sources: (i) labeled data, $\mathcal{L}=\{(Y_i,\bX_i,A_i): i=1,2,\ldots,n\}$, which are $n$ independent and identically distributed observations from the joint distribution of $(Y,\bX,A)$; and (ii) outcome-unlabeled data, $\mathcal{U}=\{(\bX_i,A_i): i=n+1,n+2,\ldots,n+N,\ N\ge 1\}$, which are $N$ independent and identically distributed observations from the marginal distribution of $(\bX,A)$. Throughout this paper, ``unlabeled'' means unlabeled with respect to the outcome $Y$; the covariates $\bX$ and the treatment assignment $A$ are observed for all subjects. Assume that (i) $\mathcal{L} \perp \mathcal{U}$, where $\perp$ represents independence; (ii) the measured pairs $(\bX,A)$ in both $\mathcal{L}$ and $\mathcal{U}$ follow the same distribution; and (iii) $\sqrt{n/N} \rightarrow \rho$ as $n\rightarrow \infty$ and $N\to \infty$, for some constant $\rho$. The outcome-unlabeled data can be of much larger size than the supervised one in various practical problems, as labeling of the outcomes is often very costly.

\subsection{Concordance-Assisted learning (CAL)}\label{sec:2.2}
The optimal treatment regime maximizing value function $E\left[Y^{*}\{d(\bX)\}\right]=E\left\{Y^{*}(1)-Y^{*}(0)\right\}$\\
$\cdot d(\bX)]+E\left\{Y^{*}(0)\right\}$ tends to assign a subject to treatment 1 if his or her $Y^{*}(1)-Y^{*}(0)>0$ and 0 otherwise, where $Y^{*}(1)-Y^{*}(0)$ is the benefit for subjects receiving treatment 1 versus treatment 0. From the view of this perspective, it is also a natural idea that, for any two subjects $i$ and $j$,
an optimal treatment regime tends to allocate subject $i$ to treatment 1 with greater chance compared to subject $j$, if $Y_{i}^{*}(1)-Y_{i}^{*}(0)>Y_{j}^{*}(1)-Y_{j}^{*}(0)$, i.e. $\beta^{\prime} \bX_{i}>\beta^{\prime} \bX_{j}$ under the linear decision rules. 
The index $\beta'\bX$ is called the prescriptive index, with the larger value of it, the subject tends to gain more benefits if assigned to treatment $1.$ Motivated by that intuition, \cite{fan2017concordance} proposed concordance-assisted learning (CAL) for estimating the optimal treatment regime in two steps. First, the estimate of $\beta$ (denoted by $\hat{\beta}$) is obtained by maximizing the concordance function
\begin{align}\label{CAL-FAN}
C(\beta)&\equiv E\left(\left[\left\{Y_{i}^{*}(1)-Y_{i}^{*}(0)\right\}-\left\{Y_{j}^{*}(1)-Y_{j}^{*}(0)\right\}\right] I\left(\beta^{\prime} \bX_{i}>\beta^{\prime} \bX_{j}\right)\right) \nonumber \\
&=E\left[\left\{D\left(\mathbf{X}_{i}\right)-D\left(\mathbf{X}_{j}\right)\right\} I\left(\beta^{\prime} \mathbf{X}_{i}>\beta^{\prime} \mathbf{X}_{j}\right)\right]
\end{align}
where $D\left(\mathbf{X}_{i}\right) \equiv E\left(Y_{i} \mid A_{i}=1, \mathbf{X}_{i}\right)-E\left(Y_{i} \mid A_{i}=0, \mathbf{X}_{i}\right)$, with the constraint $\|\beta\|=\left(\beta' \beta\right)^{1 / 2}=1 .$  Second, the estimate of $c$ (denoted by $\hat{c}$) is found through maximizing the value function $V_a \left(\hat{\beta}, c\right)=E\left[Y^{*}\left\{I\left(\hat{\beta}^{\prime} \bX \geqslant c\right)\right\}\right] .$ The optimal linear decision rule under CAL is $\hat{d}_{opt} (\bX)=I\left(\hat{\beta}' \bX \geqslant \hat{c}\right) .$ It has been shown in \cite{fan2017concordance} that the optimal treatment regime under CAL coincides with that obtained by directly maximizing the value function $\hat{V}_{a}(\beta, c)$ with respect to $\beta$ and $c$ under the constraint $\left\|\left(c,\beta^{\prime}\right)^{\prime}\right\|=1$.

\section{Semi-supervised learning with known propensity score}\label{section3}
In this section, we introduce the proposed SS-CAL to derive optimal treatment regimes utilizing both the labeled and outcome-unlabeled data, with known propensity score $\pi(\mathbf X)$ as in randomized clinical trials. The proposed estimation of OPT primarily involves a flexible nonparametric imputation by kernel smoothing which works well even when the number of covariates is large; and a follow-up estimation for optimal treatment regime based on concordance-assisted learning, including optimization of the estimated concordance function up to a threshold and finding the optimal threshold to maximize the inverse propensity score weighted (IPSW) estimator of the value function.

Denote
 \[
    V(\bX,Y,A, \nu, \pi)=\frac{\{Y-\nu(\mathbf{X})\}\{A-\pi(\mathbf{X})\}}{\pi(\mathbf{X})\{1-\pi(\mathbf{X})\}}.
 \]
Following the argument in \cite{murphy2003optimal}, we have
\begin{eqnarray}\label{m-eq:2.4}
    E\left[V(\bX,Y,A, \nu, \pi) \bigg| \bX\right]=E(Y \mid \bX, A=1)-E(Y \mid \bX, A=0)=D(\bX),
\end{eqnarray}
where $\nu(\mathbf{X})$ is an arbitrary function of $\mathbf{X}$, here we let  $\nu(\mathbf{X})\equiv \nu(\mathbf{X}, \theta)$ be a posited parametric model for $\mu(\mathbf{X}) \equiv E(Y \mid \mathbf{X}, A=0)$, such as a constant model or linear model.  

In general,  if the outcome $Y$ in $\mathcal{U}$ were actually observed, then one would simply fit the working model given in \eqref{m-eq:2.4}
and subsequently maximize the concordance function associated with the entire
dataset in $\mathcal{E} =\mathcal{L} \cup \mathcal{U}$ to estimate $\beta$. 
In order to derive the optimal treatment regime, it is essential to calculate the concordance function using equations \eqref{CAL-FAN} and \eqref{m-eq:2.4} as a primary step.
However, under the SS data framework, the function $V(\bX, Y, A, \nu, \pi)$ is only computable for labeled data $\mathcal{L}$. To utilize the outcome-unlabeled data $\mathcal{U}$, an intuitive idea is to impute the value of $V(\bX, Y, A, \nu, \pi)$ based on its conditional mean given $X$ that can be learned from the labeled data $\mathcal{L}$.
Clearly, the imputation is critical. Inaccurate imputation would result in biased estimates, while inadequate imputation leads to a loss of efficiency. However, in 
practical applications, the dimensions of covariates $\bX$ may be relatively high. There are many works in statistical and economic literature on dimension reduction, predominantly within the framework of the conditional distribution or conditional mean model (\cite{lin2018robust}). We employ the dimensional reduction technique along with a kernel method to smooth the estimate of functions with missing values over $\beta'\bX$ (the one-dimensional linear combination of possibly high-dimensional $\bX$), thereby leveraging the outcome-unlabeled data $\mathcal{U}$ to aid estimation of the optimal treatment regimes.

For linear treatment decisions, we estimate the conditional expectation of $V(\bX, Y, A, \nu, \pi)$ based on a single index $S(\bX)=\beta^{\prime} \mathbf{X}$, specifically, we consider
{\footnotesize
\begin{equation}\label{condi-E}
E[ E\{V(\bX,Y,A, \nu, \pi) \mid \bX \}]=E[E\{V(\bX,Y,A, \nu, \pi) \mid S(\boldsymbol{X})\}]=E\left[E\left\{V(\bX,Y,A, \nu, \pi) \mid \beta^T \bX \right\}\right].
\end{equation}}
Importantly, \eqref{condi-E} does not claim that
$\mathbb{E}\{V\mid \bX\}=\mathbb{E}\{V\mid S(\bX)\}$.
Rather, it motivates the single-index construction because the concordance objective
depends on $(\bX_i,\bX_j)$ only through $(S(\bX_i),S(\bX_j))$ via the indicator
$I(\beta^\top \bX_i>\beta^\top \bX_j)$, so the target can be rewritten using conditional
expectations given $(S(\bX_i),S(\bX_j))$.
The advantages of this construction and structure are as follows.
Firstly, it leverages the concept of conditional expectation and simplifies the estimation of $E[V(\bX, Y, A, \nu, \pi)|\bX]$ to the estimation of $E\left\{V(\bX, Y, A, \nu, \pi) \mid \beta'\bX \right\}$, allowing for the utilization of outcome-unlabeled information $S(\bX)$.
Secondly, it efficiently handles the estimation challenges caused by high-dimensional outcome-unlabeled data, by employing a single index representation and the projection of high-dimensional $X$ to the one-dimensional $\beta'\bX$. In contrast to the usual kernel smoothing method of letting $S(\bX) = \bX$, this approach only requires estimating the single index function $E\left\{V(\bX, Y, A, \nu, \pi) \mid \beta' \bX \right\}$, enables us to reduce the dimension of the regressor from $p$ to 1 and thus greatly reducing computational complexity and estimation difficulty.
The third advantage lies in the utilization of the linear part $\beta' \bX$ information in the treatment decision $d(\bX) = I(\beta' \bX \geq c)$, which enhances the utilization of the decision function and increases the information regarding treatment decisions in the estimation function.
The fourth advantage pertains to SS learning on outcome-unlabeled data, utilizing the information of $\beta' \bX$ without introducing new unknown parameters into the estimation process.

By equations \eqref{m-eq:2.4} and \eqref{condi-E} we have
\begin{equation}\label{EV}
\begin{aligned}
E[ V(\bX,Y,A, \nu, \pi) |\beta^{\prime} \mathbf{X}]&=E\Big \{ E\Big [ \frac{\{Y-\nu(\mathbf{X})\}\{A-\pi(\mathbf{X})\}}{\pi(\mathbf{X})\{1-\pi(\mathbf{X})\}} \Big |\bX\Big] \Big | \beta^{\prime} \mathbf{X} \Big\}\\
&=E\Big \{ E [Y|\bX,A=1]-E(Y|\bX,A=0) \big | \beta^{\prime} \mathbf{X} \Big\}\\
&=E[D(\bX)|\beta^{\prime} \mathbf{X}].
\end{aligned}
\end{equation}
Next, since $I(\beta^\top X_i>\beta^\top X_j)$ is measurable with respect to
$(S(\bX_i),S(\bX_j))=(\beta^\top \bX_i,\beta^\top \bX_j)$, we have
\begin{align}\label{EC}
  C(\beta) &=E\left\{E[  D(\bX_i)-D(\bX_j) |\beta^{\prime}\bX_i, \beta^{\prime}\bX_j] I\left(\beta^{\prime} \mathbf{X}_{i}>\beta^{\prime} \mathbf{X}_{j}\right)\right\}\nonumber\\
    &= E\left[\left\{  E[ V(\bX_i,Y_i,A_i, \nu, \pi) |\beta^{\prime} \mathbf{X}_i]  - E[ V(\bX_j,Y_j,A_j, \nu, \pi) |\beta^{\prime} \mathbf{X}_j] \right\} I\left(\beta^{\prime} \mathbf{X}_{i}>\beta^{\prime} \mathbf{X}_{j}\right)\right]\nonumber\\
    &=E\left\{\left[  m_\beta(\beta^{\prime} \mathbf{X}_i,\nu, \pi)  -m_\beta(\beta^{\prime} \mathbf{X}_j,\nu, \pi) \right] I\left(\beta^{\prime} \mathbf{X}_{i}>\beta^{\prime} \mathbf{X}_{j}\right)\right\},
\end{align}
where $m_\beta(s, \nu, \pi)=E\{ V(\bX,Y,A, \nu, \pi) \mid \beta^{\prime} \mathbf{X}=s\}$; the subscript emphasizes that the conditional mean is defined with respect to the current index $S_\beta=\beta^{\prime} \mathbf{X}$.
Therefore, we can derive the SS estimator of the concordance function $C(\beta)$ based on the conditional mean $m_\beta(\beta^{\prime} \mathbf{X},\nu, \pi)$. 
If we could propose a suitable estimator of $m_\beta(\beta^{\prime} \mathbf{X},\nu, \pi)$ for outcome-unlabeled data under a given parameter $\beta$ of interest, we can therefore obtain a reliable estimator of $C(\beta)$ by \eqref{EC} associated with both data sets, and the optimal treatment regime can also be derived.

In order to obtain an estimator of $C(\beta)$, we need to pre-estimate $m_\beta(S_\beta,\nu,\pi)$ for a given $\beta$ because  $m_\beta(S_\beta,\nu,\pi)$ is unknown for outcome-unlabeled data.
One way to avoid model misspecification is to estimate $m_\beta(S_\beta,\nu,\pi)$ using a nonparametric regression so that no specific form is assumed.
The law of iterated expectation in equations \eqref{condi-E} and \eqref{EC} enables us to utilize the single index for imputation without making specific assumptions about data. Hence, from \eqref{EV},  we propose a semiparametric approach for estimating $D(\bX)$ by Nadaraya-Watson (NW) kernel estimator 
\begin{eqnarray}\label{hatD}
   \hat{m}_{\beta}(\beta^{\prime}\bX_l,\nu,\pi)=\frac{n^{-1} \sum_{i=1}^n K_h(\beta^{\prime} \bX_i-\beta^{\prime} \bX_l)V(\bX_i,Y_i,A_i,\nu,\pi)}{n^{-1} \sum_{i=1}^nK_h(\beta^{\prime} \bX_i-\beta^{\prime} \bX_l)}, \quad l=n+1,...,n+N,
\end{eqnarray}
and $K_h(u-v) = h^{-1}K\{(u -v)/h\}$ with $K$: $\mathbb{R}  \rightarrow \mathbb{R}$ being some suitable kernel
function and $h = h(n) > 0$ being the bandwidth parameter that satisfies often $h\to 0$ and $nh\to \infty$ as $n\to \infty$. 
To ensure theoretical rigor, the theoretical analysis below uses the trimmed version of the NW kernel estimator to avoid problems that might be caused by the denominator in (\ref{hatD}) being too small. With a slight abuse of notation, we still denote the trimmed estimator by $\widehat m_\beta$. More precisely, equation \eqref{hatD} is replaced in the theoretical analysis by 
$\tilde{m}_{\beta}(\beta^{\prime}\bX_l,\nu,\pi)=\frac{n^{-1}\sum_{i=1}^n K_h(\beta^{\prime} \bX_i-\beta^{\prime} \bX_l)V(\bX_i, Y_i, A_i,\nu,\pi)}{\Tilde f_{n,\beta}(\beta^{\prime}\bX_l)}, \quad l=n+1,...,n+N$, where $\hat f_{n,\beta}(\beta^{\prime}\bX_l)=n^{-1}\sum_{i=1}^n K_h(\beta^{\prime} \bX_i-\beta^{\prime}\bX_l)$, $\Tilde f_{n,\beta}(\beta^{\prime}\bX_l) = \max \big\{\hat f_{n,\beta}(\beta^{\prime}\bX_l), w_n\big\}$ and $w_n$ is a decreasing positive constant sequence which will be described in Section \ref{section5}.
In fact, we need an undersmooth condition of $h$ for the estimation of this index model $m_\beta(S_\beta,\nu,\pi)=E\{ V(\bX, Y, A, \nu, \pi) \mid S_\beta\}$, which is given as Condition 1 in Section \ref{section5}.

We could posit a parametric model (e.g. linear model) $\nu(\mathbf{X},\theta)$ for $\mu(\mathbf{X}) \equiv E(Y \mid \mathbf{X}, A=0)$ to the labeled data and get the estimator of $\theta$ (denoted by  $\hat{\theta}$, e.g. least square estimator) and $\hat{\nu}\equiv \nu(\mathbf{X},\hat{\theta}).$ With the NW kernel estimator as defined in \eqref{hatD}, we propose a SS estimator for the concordance function $C(\beta)$, that is
\begin{eqnarray}\label{m-eq:2.6}
    \hat{C}_{\lambda}(\beta,\hat{\theta})
   &\equiv \frac{\lambda}{n(n-1)} \sum_{1\le i \neq j\le n}  (V(\bX_i,Y_i,A_i, \hat{\nu}, \pi) -V(\bX_j,Y_j,A_j, \hat{\nu}, \pi) ) I\left(\beta^{\prime} \mathbf{X}_{i}>\beta^{\prime} \mathbf{X}_{j}\right)\nonumber\\
   &+\frac{1-\lambda}{N(N-1)} \sum_{n+1\le i \neq j \le n+N}\{  \widehat{m}_{\beta}(\beta^{\prime}\bX_i,\hat{\nu},\pi)-\widehat{m}_{\beta}(\beta^{\prime}\bX_j,\hat{\nu},\pi)\} I\left(\beta^{\prime} \mathbf{X}_{i}>\beta^{\prime} \mathbf{X}_{j}\right),
\end{eqnarray}
where $\lambda \in [0,1]$ is a weight to be decided. 
Here the estimator $\hat{C}_{\lambda}(\beta, \hat{\theta})$ incorporates
both the concordance information in the labeled data with weight $\lambda$ and that in the outcome-unlabeled data with weight $1-\lambda$ based on the imputed values obtained by suitable training on the labeled dataset. 
If we set $\lambda=1$, the induced estimator degenerates to the fully supervised estimator by \cite{fan2017concordance}, which utilizes the labeled data only, and is shown to lose efficiency in the SS setting compared to our proposed method, both in the theoretical properties and in the simulation results.

Denote $\hat{\beta}_{\lambda}=\arg \max _{\|\boldsymbol{\beta}\|=1} \hat{C}_{\lambda}(\beta, \hat{\theta})$ and an estimator of $c^{*}$ is given by $\hat{c}_{\lambda}=\arg \max _{c} \hat{V}_{a}(\hat{\beta}_{\lambda}, c)$.

We also propose another estimator of the concordance function utilizing the concordance among both the labeled and outcome-unlabeled data:
{\small
\begin{eqnarray}\label{m-eq:2.7}
    \hat{C}_{pl}(\beta,\hat{\theta}) \equiv \frac{1}{(n+N)(n+N-1)} \sum_{1\le i \neq j \le n+N}\{  \widehat{m}_{\beta}(\beta^{\prime}\bX_i,\hat{\nu},\pi)-\widehat{m}_{\beta}(\beta^{\prime}\bX_j, \hat{\nu},\pi)\} I\left(\beta^{\prime} \mathbf{X}_{i}>\beta^{\prime} \mathbf{X}_{j}\right).
\end{eqnarray}
}
This estimator $ \hat{C}_{pl}(\beta,\hat{\theta})$ is based on the average of predicted treatment difference among all subjects from the labeled and outcome-unlabeled samples $\mathcal{E}=\mathcal{L} \cup \mathcal{U}$.
For convenience, we call this estimator the “pool estimator”, because it utilizes all observed samples, just like pooling all the samples together.

Denote $\hat{\beta}_{pl}=\arg \max_{\|\boldsymbol{\beta}\|=1} 
\hat{C}_{pl}(\beta,\hat{\theta})$ and an estimator of $c^{*}$ is given by $\hat{c}_{pl}=\arg \max _{c} \hat{V}_{a}(\hat{\beta}_{pl}, c)$.

The optimization process for the estimated concordance function involves a kernel estimator where $\beta$ is unknown. To simplify this, we adopt a two-step iterative algorithm for implementation. Using the labeled data $\mathcal{L}$, we easily derive a reliable initial estimator $\hat{\bbeta}_{init}$, and $\hat{\btheta}$, which minimizes $\frac{1}{n} \sum_{i=1}^n (1-A_i)[Y_i-\nu(\bX_i,\btheta)]^2$. First, we employ the current estimator, $\hat{\bbeta}$, to compute values for $\widehat{m}_{\beta}(\beta^{\prime}\bX_i,\nu(\mathbf{X},\hat{\theta}),\pi)$. Second, we maximize the estimated concordance function using these computed values $\widehat{m}_{\beta}(\beta^{\prime}\bX_i,\nu(\mathbf{X},\hat{\theta}),\pi)$ to update the estimator for $\bbeta$. This process iterates between these two steps until convergence is achieved.

\section{Semi-supervised learning with unknown propensity score}\label{section4}
In this section, we assume the propensity score $\pi (\bX)$ is unknown, like in observational studies, and study how to derive optimal treatment regimes with unknown propensity score under SS setting.
Denote $\pi(\mathbf{X}, \alpha)$ as the posited propensity score model; a parametric model (e.g., a logistic regression model) is usually assumed for the propensity score.  Let $\hat{\alpha}$ denote the estimator of the parameters $\alpha$ obtained from the labeled sample. Although $\pi(\mathbf{X}, \hat{\alpha})$ may not consistently estimate the true propensity score $\pi(\mathbf{X})$ when the propensity score model is misspecified, it can be shown that $\hat{\alpha}$ converges almost surely to a deterministic vector of parameters $\alpha^{*}$ under mild conditions. If the propensity score model is misspecified,
$\hat{C}_{\lambda}(\beta,\hat{\theta})$ and $ \hat{C}_{pl}(\beta,\hat{\theta})$ are not consistent estimators of the concordance function.

Therefore, to improve robustness against propensity score misspecification,
a doubly robust estimation method is developed for the class of monotonic index models: $D(\mathbf{X})=Q\left({\beta^*}^{\prime} \mathbf{X}\right)$, where $Q(\cdot)$ is a non-constant and increasing function of the index ${\beta^*}^{\prime} \mathbf{X}$.

Define
$$
    G^{\mathrm{DR}}_{ij}(\theta, \alpha)=V(\bX_i,Y_i, A_i, \nu, \pi)\frac{A_j}{\pi (\bX_j,\alpha)}=\frac{\left\{Y_{i}-\nu\left(\mathbf{X}_{i}, \theta\right)\right\}\left\{A_{i}-\pi\left(\mathbf{X}_{i}, \alpha\right)\right\} A_{j}}{\pi\left(\mathbf{X}_{i}, \alpha\right)\left\{1-\pi\left(\mathbf{X}_{i}, \alpha\right)\right\} \pi\left(\mathbf{X}_{j}, \alpha\right)},
 $$
 $$
     G^{\mathrm{DR}}_{ji}(\theta, \alpha)=V(\bX_j,Y_j, A_j, \nu, \pi)\frac{A_i}{\pi (\bX_i,\alpha)}=\frac{\left\{Y_{j}-\nu\left(\mathbf{X}_{j}, \theta \right)\right\}\left\{A_{j}-\pi\left(\mathbf{X}_{j}, \alpha\right)\right\} A_{i}}{\pi\left(\mathbf{X}_{j}, \alpha\right)\left\{1-\pi\left(\mathbf{X}_{j}, \alpha\right)\right\} \pi\left(\mathbf{X}_{i}, \alpha\right)},
 $$
 $$
\Lambda_{i j}^{\mathrm{DR}}(\theta, \alpha)=G^{\mathrm{DR}}_{ij}(\theta, \alpha)- G^{\mathrm{DR}}_{ji}(\theta, \alpha)
$$
where $\nu(\mathbf{X}, \theta)$ is a posited parametric model for $\mu(\mathbf{X})$.

Recall that $\mu(\mathbf{X})=E(Y \mid \mathbf{X}, A=0)$ and $E(A_j|\mathbf{X})=\pi (\mathbf{X}_j)$, we have
$$
E\left[\frac{\{Y-\mu(\mathbf{X})\}\{A-\pi(\mathbf{X}, \alpha)\}}{\pi(\mathbf{X}, \alpha)\{1-\pi(\mathbf{X}, \alpha)\}} \bigg| \mathbf{X}\right]=Q\left({\beta^*}^{\prime} \mathbf{X}\right) \frac{\pi(\mathbf{X})}{\pi(\mathbf{X}, \alpha)}.
$$
In summary, if the propensity score model is correctly specified,
$$
E\left\{\Lambda_{i j}^{\mathrm{DR}}(\theta, \alpha) I\left(\beta^{\prime} \mathbf{X}_{i}>\beta^{\prime} \mathbf{X}_{j}\right)\right\}=E\left[\left\{Q\left({\beta^*}^{\prime} \mathbf{X}_{i}\right)-Q\left({\beta^*}^{\prime} \mathbf{X}_{j}\right)\right\} I\left(\beta^{\prime} \mathbf{X}_{i}>\beta^{\prime} \mathbf{X}_{j}\right)\right].
$$
Whereas, if the baseline mean model $\nu(\mathbf{X}, \theta)$ is correctly specified,
{\small
$$
E\left\{\Lambda_{i j}^{\mathrm{DR}}(\theta, \alpha) I\left(\beta^{\prime} \mathbf{X}_{i}>\beta^{\prime} \mathbf{X}_{j}\right)\right\}=E\left[\frac{\pi\left(\mathbf{X}_{i}\right) \pi\left(\mathbf{X}_{j}\right)}{\pi\left(\mathbf{X}_{i}, \alpha\right) \pi\left(\mathbf{X}_{j}, \alpha\right)}\left\{Q\left({\beta^*}^{\prime} \mathbf{X}_{i}\right)
-Q\left({\beta^*}^{\prime} \mathbf{X}_{j}\right)\right\} I\left(\beta^{\prime} \mathbf{X}_{i}>\beta^{\prime} \mathbf{X}_{j}\right)\right].
$$
}
Under either case, the maximizer of $E\left\{\Lambda_{i j}^{\mathrm{DR}}(\theta, \alpha) I\left(\beta^{\prime} \mathbf{X}_{i}>\beta^{\prime} \mathbf{X}_{j}\right)\right\}$ is $\beta=\beta^{*}$ as $D(\mathbf{X})=Q\left({\beta^*}^{\prime} \mathbf{X}\right)$ representing the class of monotonic index models.
We define the doubly robust concordance function
$$
\begin{aligned}
&
C^{\mathrm{DR}}(\beta, \theta,\alpha)\equiv E\{\Lambda_{i j}^{\mathrm{DR}}(\theta, \alpha) I\left(\beta^{\prime} \mathbf{X}_{i}>\beta^{\prime} \mathbf{X}_{j}\right)\}\\
&=E[ E \{\Lambda_{i j}^{\mathrm{DR}}(\theta, \alpha)|\beta^{\prime} \mathbf{X}_{i}, \beta^{\prime} \mathbf{X}_{j} \} I(\beta^{\prime} \mathbf{X}_{i}>\beta^{\prime} \mathbf{X}_{j})]\\
&= E\left\{\left[E\left\{G_{i j}^{\mathrm{DR}}(\theta, \alpha) \mid \beta^{\prime} \mathbf{X}_{i}, \beta^{\prime} \mathbf{X}_{j}\right\}\right.\right.
\left.\left.-E\left\{G_{j i}^{\mathrm{DR}}(\theta, \alpha) \mid \beta^{\prime} \mathbf{X}_{i}, \beta^{\prime} \mathbf{X}_{j}\right\}\right] I\left(\beta^{\prime} \mathbf{X}_{i}>\beta^{\prime} \mathbf{X}_{j}\right)\right\} .
\end{aligned}
$$
For the outcome-unlabeled data, we could estimate $
E\left\{G_{i j}^{\mathrm{DR}}(\theta, \alpha) \mid \beta^{\prime} \mathbf{X}_{i},\beta^{\prime} \mathbf{X}_{j}\right\}
$
by a nonparametric kernel method based on the labeled data, utilizing the multidimensional smoother, product kernel, that is,
\begin{equation}\label{hat_DR}
\hat{G}_{i j}^{\mathrm{DR}}(\theta, \alpha)=\frac{1/n^2\sum_{l=1}^{n} \sum_{m=1}^{n} K_{h_{1}}\left(\beta^{\prime} \mathbf{X}_{l}-\beta^{\prime} \mathbf{X}_{i}\right) K_{h_{2}}\left(\beta^{\prime} \mathbf{X}_{m}-\beta^{\prime} \mathbf{X}_{j}\right) G_{l m}^{\mathrm{DR}}(\theta, \alpha)}{1/n^2\sum_{l=1}^{n} \sum_{m=1}^{n} K_{h_{1}}\left(\beta^{\prime} \mathbf{X}_{l}-\beta^{\prime} \mathbf{X}_{i}\right) K_{h_{2}}\left(\beta^{\prime} \mathbf{X}_{m}-\beta^{\prime} \mathbf{X}_{j}\right)},
\end{equation}
where $K_{h_\ell}(u-v)=h_\ell^{-1}K\{(u-v)/h_\ell\}$ for $\ell=1,2$, with $K:\mathbb{R}\rightarrow\mathbb{R}$ being a suitable kernel function and $h_1=h_1(n)>0$, $h_2=h_2(n)>0$ being the bandwidths.
Analogously, for the theoretical analysis of the DR estimator we use the trimmed product-kernel version and, with a slight abuse of notation, still write it as $\widehat G_{ij}^{\mathrm{DR}}$. More precisely, equation \eqref{hat_DR} is replaced in the theoretical analysis by
$$\tilde{G}_{i j}^{\mathrm{DR}}(\theta, \alpha)=\frac{1/n^2\sum_{l=1}^{n} \sum_{m=1}^{n} K_{h_{1}}\left(\beta^{\prime} \mathbf{X}_{l}-\beta^{\prime} \mathbf{X}_{i}\right) K_{h_{2}}\left(\beta^{\prime} \mathbf{X}_{m}-\beta^{\prime} \mathbf{X}_{j}\right) G_{l m}^{\mathrm{DR}}(\theta, \alpha)}{\Tilde f_{n,\beta}^{(2)}(\beta^{\prime}\bX_i, \beta^{\prime}\bX_j)},$$
where
$\hat f_{n,\beta}^{(2)}(\beta^{\prime}\bX_i,\beta^{\prime}\bX_j)=1/n^2\sum_{l=1}^{n} \sum_{m=1}^{n} K_{h_{1}}\left(\beta^{\prime} \mathbf{X}_{l}-\beta^{\prime} \mathbf{X}_{i}\right) K_{h_{2}}\left(\beta^{\prime} \mathbf{X}_{m}-\beta^{\prime} \mathbf{X}_{j}\right)$, $\Tilde f_{n,\beta}^{(2)}(\beta^{\prime}\bX_i, \beta^{\prime}\bX_j) = \max \big\{\hat f_{n,\beta}^{(2)}(\beta^{\prime}\bX_i,\beta^{\prime}\bX_j), w_n\big\}$ and $w_n$ is a decreasing positive constant sequence which will be described in Section \ref{section5}.

With $\widehat{G}^{DR}(\cdot)$ as defined in \eqref{hat_DR}, we update the SS estimator of the doubly robust concordance function by
{\small
\begin{eqnarray}\label{m-eq:2.8}
    \hat{C}_{\lambda}^{\mathrm{DR}}(\beta, \hat{\theta}, \hat{\alpha})
   &\equiv&  \frac{\lambda}{n(n-1)} \sum_{1\le i \neq j\le n} \Lambda^{\mathrm{DR}}_{i j}(\hat{\theta},\hat{\alpha}) I\left(\beta^{\prime} \mathbf{X}_{i}>\beta^{\prime} \mathbf{X}_{j}\right)\nonumber\\
   &&+\frac{1-\lambda}{N(N-1)} \sum_{n+1\le i \neq j \le n+N}\{  \widehat{G}^{\mathrm{DR}}_{ij}(\hat{\theta},\hat{\alpha})-\widehat{G}^{\mathrm{DR}}_{ji}(\hat{\theta},\hat{\alpha})\} I\left(\beta^{\prime} \mathbf{X}_{i}>\beta^{\prime} \mathbf{X}_{j}\right),
\end{eqnarray}
}
where $\lambda \in [0,1]$ is a weight to be decided.
If we set $\lambda=1$, the induced estimator degenerates to the fully supervised doubly robust estimator by \cite{fan2017concordance} that utilizes the labeled data only, which is shown to lose efficiency in the SS setting compared to our proposed method both in the theoretical results and also in the simulation studies.
Denote $\hat{\beta}_{\lambda}^{\mathrm{DR}}=\arg \max _{\|\boldsymbol{\beta}\|=1} \hat{C}_{\lambda}^{\mathrm{DR}}(\beta, \hat{\theta}, \hat{\alpha})$.
An estimator of $c^{*}$ is given by $\hat{c}_{\lambda}^{\mathrm{DR}}=\arg \max _{c} \hat{V}_{a}\left(\hat{\beta}_{\lambda}^{\mathrm{DR}}, \hat{\alpha}, c\right)$.

We also propose another SS estimator for the concordance function:
{\small
\begin{eqnarray}\label{m-eq:2.9}
    \hat{C}_{pl}^{\mathrm{DR}}(\beta, \hat{\theta}, \hat{\alpha})
   \equiv \frac{1}{(n+N)(n+N-1)} \sum_{1\le i \neq j \le n+N}\{  \widehat{G}^{\mathrm{DR}}_{ij}(\hat{\theta},\hat{\alpha})-\widehat{G}^{\mathrm{DR}}_{ji}(\hat{\theta},\hat{\alpha})\} I\left(\beta^{\prime} \mathbf{X}_{i}>\beta^{\prime} \mathbf{X}_{j}\right).
\end{eqnarray}
}
This estimator of the doubly robust concordance function is based on the average of predicted treatment difference among all subjects from the labeled and outcome-unlabeled samples $\mathcal{E}=\mathcal{L} \cup \mathcal{U}$.
Denote $\hat{\beta}_{pl}^{\mathrm{DR}}=\arg \max _{\|\beta\|=1}  \hat{C}_{pl}^{\mathrm{DR}}(\beta, \hat{\theta}, \hat{\alpha})$.
The relative estimator for $c^*$ is given by
$
\hat{c}_{p l}^{\mathrm{DR}}=\arg \max _{c} \hat{V}_{a}\left(\hat{\beta}_{p l}^{\mathrm{DR}}, \hat{\alpha}, c\right).
$

\section{Properties of the proposed estimator} \label{section5}
To study the asymptotic properties of the estimators, we first introduce some notation for simplicity of presentation. Define
$
\tau\left(\beta, \mathbf{X}_{1}, \mathbf{X}_{2}\right)=\left\{D\left(\mathbf{X}_{1}\right)-D\left(\mathbf{X}_{2}\right)\right\} I\left(\beta^{\prime} \mathbf{X}_{1}>\beta^{\prime} \mathbf{X}_{2}\right)
$
and
$
\varrho(\beta, \mathbf{x})=E\{\tau(\beta, \mathbf{x}, \mathbf{X})\}+E\{\tau(\beta, \mathbf{X}, \mathbf{x})\}.
$
Let $\nabla_{m} \varrho(\beta, \mathbf{x})$ denote the $m$th derivative with respect to $\beta$ (in particular, $\nabla_1$ and $\nabla_2$ denote the first and second derivatives with respect to $\beta$), and define
$
\left|\nabla_{m}\right| \varrho(\beta, \mathbf{x})=\sum_{i_{1}+\ldots+i_{m}=m}\left|\frac{\partial^{m} \varrho(\beta, \mathbf{x})}{\partial \beta_{i_{1}} \ldots \partial \beta_{i_{m}}}\right| .
$
To establish the asymptotic results, we need the following conditions.

Condition 1. (a) The kernel function $K(\cdot): \mathbb{R} \rightarrow \mathbb{R}$ is a symmetric probability density function satisfying $\int K^2(u) d u<\infty$ and $\mu_2(K)=\int u^2 K(u) d u<$ $\infty$. 
$K(s)$ is twice continuously differentiable, and the second derivative satisfies a Lipschitz condition.
The smoothing bandwidth $h$ satisfies $n h \rightarrow \infty$ and $n h^4 \rightarrow 0$ as $n \rightarrow \infty$.

(b) Let $\mathbf{X}=\left(X_1, \ldots, X_p\right)^{\prime}$. Assume that there exists $\delta>0$ such that $\max _{1 \leq k \leq p} E\left|X_k\right|^{4+\delta}<\infty$. The labeled data and outcome-unlabeled data have the same joint distribution of $(\mathbf{X},A)$, and hence the same covariate distribution.
 
Condition 2. The propensity score $\pi(\mathbf{x})$ is known and $0<\pi(\mathbf{x})<1$ for all $\mathbf{x} \in \mathcal{X}$.

Condition 3. The estimator $\hat{\theta}$ converges almost surely to a deterministic vector of parameters $\theta^{*}$, and $n^{1 / 2} (\hat{\theta}-\theta^{*} )=O_{p}(1)$.

Condition 4. The concordance function $C(\beta)=E\left\{\tau\left(\beta, \mathbf{X}_{1}, \mathbf{X}_{2}\right)\right\}$ has a unique maximizer at $\beta=\beta^{*}=\left(\beta_{1}^{*}, \ldots, \beta_{p}^{*}\right)^{\prime}$ with $\left\|\beta^{*}\right\|=1$.

Condition 5.
(a) The support of $\mathbf{X}$ is not contained in a proper linear subspace of $\mathbb{R}^{p}$.

(b) For each $\beta\in\mathcal B$, let $f_\beta$ and $F_\beta$ denote the density and the cumulative distribution function of $S_\beta=\beta^{\prime}\bX$, respectively, and let $\mathcal{A}_\beta$ be the support of $S_\beta$. Assume that $f_\beta$ is continuous on the interior of $\mathcal{A}_\beta$, is strictly positive there, and satisfies $f_\beta(s)\to 0$ as $s\to \partial\mathcal{A}_\beta$. Let
\[
m_{0,\beta}(s,\theta^*)=E\{\psi(\bZ,\theta^*)\mid S_\beta=s\}
\]
be the oracle conditional mean used in the kernel-imputation step. There exist a decreasing positive sequence $w_n\downarrow 0$ and a positive sequence $\epsilon_n$ such that $\epsilon_n\asymp c_n$, $\epsilon_n/w_n\to 0$, and uniformly for $\beta\in\mathcal{B}$, $\sqrt{n}\,P\{f_\beta(S_\beta)\le w_n+\epsilon_n\}\to 0$,
$\sqrt{n}\,E\big[|m_{0,\beta}(S_\beta,\theta^*)|\,I\{f_\beta(S_\beta)\le w_n+\epsilon_n\}\big]\to 0$,
$c_n^2/w_n^2=o(n^{-1/2}),\qquad w_n^2\sqrt{n}\to\infty$,
where $c_{n}=h^{2}+\left[\log h^{-1} /(n h)\right]^{1 / 2}$.

(c) $E\left\{D(\mathbf{X})^{2}\right\}<\infty$ and
$
E\left(\left[\frac{\left\{Y-\nu\left(\mathbf{X}, \theta^{*}\right)\right\}\{A-\pi(\mathbf{X})\}}{\pi(\mathbf{X})\{1-\pi(\mathbf{X})\}}\right]^{2}\right)<\infty.
$

Condition 6. (a) The function $\varrho(\beta, \mathbf{x})$ is twice differentiable with respect to $\beta$.

(b) There is an integrable function $\Upsilon(\mathbf{x})$ such that, for any $\mathbf{x} \in \mathcal{X}$ and $\beta_{1}$ and $\beta_{2}$ with $\left\|\beta_{1}\right\|=\left\|\beta_{2}\right\|=1,\left\|\nabla_{2} \varrho\left(\beta_{1}, \mathbf{x}\right)-\nabla_{2} \varrho\left(\beta_{2}, \mathbf{x}\right)\right\|<\Upsilon(\mathbf{x})\left\|\beta_{1}-\beta_{2}\right\|$.

(c) $E\left\{\left|\nabla_{1} \varrho\left(\beta^{*}, \mathbf{X}\right)\right|^{2}\right\}<\infty$, $E\left\{\|\nabla_{2}\varrho(\beta^{*},\mathbf X)\|_F\right\}<\infty$, and $E\left\{\nabla_{2} \varrho\left(\beta^{*}, \mathbf{X}\right)\right\}$ is negative definite.

(d) With $S_\beta=\beta^{\prime}\mathbf X$, define
$
\mathcal W(\beta)=\psi(\mathbf Z,\theta^*)-m_{0,\beta}(S_\beta,\theta^*),
\phi(\beta)=2F_\beta(S_\beta)-1.
$
The map $\beta\mapsto\mathcal W(\beta)\phi(\beta)$ admits a second-order expansion in a neighborhood of $\beta^*$ with an integrable envelope, differentiation may be interchanged with expectation, and
$
q(\mathbf Z)=\left.\nabla_\beta\{\mathcal W(\beta)\phi(\beta)\}\right|_{\beta=\beta^*}
$
exists and satisfies $E\{\|q(\mathbf Z)\|^2\}<\infty$.

Throughout the asymptotic analysis, all derivatives with respect to $\beta$ and all inverses involving $V$ or $V^{\mathrm{DR}}$ are understood in a local $(p-1)$-dimensional parametrization of the unit sphere $\{\beta:\|\beta\|=1\}$, or equivalently on the tangent space at $\beta^*$.

Condition 7. The value function $V_a\left(\beta^{*}, c\right)=E\left[Y^{*}\left\{I\left(\beta^{* \prime} \mathbf{X} \geqslant c\right)\right\}\right]$ has a unique maximizer at $c=c^{*}$. In addition, there is a neighbourhood of $c^{*}$ and a constant $\kappa>0$ such that $V_a\left(\beta^{*}, c\right)-$ $V_a\left(\beta^{*}, c^{*}\right) \leqslant-\kappa\left(c-c^{*}\right)^{2}$ for every $c$ in this neighbourhood.

Condition 1(a) is a commonly used condition for kernel estimation, which requires undersmoothing of bandwidths and is standard to obtain consistency of semiparametric estimators (\cite{tsiatis2006semiparametric}). In Condition 1(b), we assume finite higher moments to remove the effects of estimating $\beta$, which is slightly stronger than typical moment requirements for asymptotic normality. Similar techniques are adopted in \cite{hu2012semiparametric}. Conditions 2-7 are general conditions that are also used in \cite{fan2017concordance}. Condition 2 is assumed for simplicity, which can be extended to the situation when the propensity score model is correctly specified. Under mild conditions, Condition 3 usually holds for the least squares estimator. Conditions 4 and 5 are assumed to establish the consistency of $\hat{\beta}$. Condition 6 is assumed to ensure the asymptotic normality of $\hat{\beta}$ and to justify the first-order kernel-imputation correction term. Conditions 5 and 6 are often used to establish the large sample properties of maximum rank correlation estimators (e.g. \cite{sherman1993limiting} and \cite{cavanagh1998rank}). The revised Condition 5(b) is the trimmed analogue of the boundary control used in the proof of the trimmed semi-supervised kernel estimator: it isolates the low-density region through $w_n$ and directly requires the tail contribution of that region to be $o(n^{-1/2})$. This is exactly the regime needed to justify replacing the raw kernel estimator by the trimmed one in the concordance objective. Condition 7 is assumed to establish the consistency and convergence rate of $\hat{c}$. 

Denote $\bZ_k=(\bX_k,Y_k,A_k)$, $S_k=S_{\beta,k}=\beta^\top\bX_k$, $F_\beta(s)=P(\beta^\top\bX\le s)$, $\phi_k(\beta)=2F_\beta(S_{\beta,k})-1$, $\psi(\bZ_k,\theta^*)=V(X_k,Y_k,A_k,\nu(\theta^*),\pi)$, $m_{0,\beta}(s, \theta^*) = E\left\{V(\bX_k,Y_k,A_k, \nu(\theta^*), \pi)\mid S_{\beta,k}= s\right\}$, and $\mathcal{W}_k (\beta) =  \psi(\bZ_k,\theta^*) - m_{0,\beta}(S_{\beta,k}, \theta^*)$.
   
\begin{theorem}\label{theorem1}
Under conditions $1-6$, when $n \rightarrow \infty$, $n \leq N$, $\sqrt{n/N} \rightarrow \rho$, we have

(a) $\left\|\hat{\beta}_{\lambda}-\beta^{*}\right\| \rightarrow 0$ almost surely;

(b) $\sqrt{n}\left(\hat{\beta}_{\lambda}-\beta^{*}\right) \rightarrow N(0, \mathbf{\Sigma}_{\lambda})$ in distribution;

(c) $\left\|\hat{\beta}_{pl}-\beta^{*}\right\| \rightarrow 0$ almost surely;

(d) $\sqrt{n}\left(\hat{\beta}_{pl}-\beta^{*}\right) \rightarrow N(0, \mathbf{\Sigma}_{pl} )$ in distribution;

where $
\boldsymbol{\Sigma}_\lambda=
V^{-1}\Omega_\lambda V^{-1},
\boldsymbol{\Sigma}_{pl}=V^{-1}\Omega_{pl}V^{-1},
$
with
$
\Omega_\lambda=
\{\lambda^2+(1-\lambda)^2\rho^2\}\Delta
+(1-\lambda)^2\Delta_2
+\lambda(1-\lambda)(\Gamma+\Gamma^{\prime}),
$
$
\Omega_{pl}=\frac{\rho^2}{1+\rho^2}\Delta
+\Delta_2
+\frac{\rho^2}{1+\rho^2}(\Gamma+\Gamma^{\prime}).
$
Here
$
2V=E\{\nabla_2\varrho(\beta^*,\mathbf X)\},
A_k=\nabla_1\varrho(\beta^*,\mathbf X_k),
q_k=\left.\nabla_\beta\{\mathcal W_k(\beta)\phi_k(\beta)\}\right|_{\beta=\beta^*},
$
$
\Delta=E(A_kA_k^{\prime}),
\Delta_2=E(q_kq_k^{\prime}),
\Gamma=E(A_kq_k^{\prime}).
$
Here $V^{-1}\Delta V^{-1}$ is the asymptotic variance for the fully supervised estimator by \cite{fan2017concordance}. If $\Gamma=0$ (the cross covariance between the oracle rank-expansion term and the kernel-imputation correction term vanishes) and the imputation-correction variance is negligible, e.g. $\Delta_2=0$, then the simple rule $\lambda=\rho^2/(1+\rho^2)=n/(n+N)$ is recovered.
\end{theorem}

Because the cross-covariance term is retained in $\Omega_\lambda$, there is in general no universal closed-form weight that minimizes $\boldsymbol{\Sigma}_\lambda$ in the Loewner order. In practice, one may choose $\lambda$ by a one-dimensional grid search that minimizes a scalar summary, such as $\operatorname{tr}(\widehat{\boldsymbol{\Sigma}}_\lambda)$, using the plug-in or perturbation-resampling variance estimate. The simple choice $\lambda=n/(n+N)$ may still be used as a computationally convenient default.
Since the asymptotic variance matrix $\Sigma_\lambda$ has a very complicated form, the direct estimation of $\Sigma_\lambda$ may be difficult due to the complicated functional forms. Instead, we propose a perturbation-based bootstrap method in Section S1 and additional theoretical results for constant $c^*$ and the value function are relegated to 
Section 2 in the supplementary material.

\begin{theorem}\label{theorem4}
Assume that either the propensity score model $\pi(\mathbf{X},\alpha)$ or the baseline mean model $\nu(\mathbf{X},\theta)$ is correctly specified and that
\[
D(\mathbf{X})=E\{Y^*(1)-Y^*(0)\mid \mathbf{X}\}=Q({\beta^*}^{\prime}\mathbf{X}),
\]
where $Q(\cdot)$ is a non-constant increasing function. Let $\eta=(\theta^\prime,\alpha^\prime)^\prime$ and assume that $\hat\eta=(\hat\theta^\prime,\hat\alpha^\prime)^\prime$ is estimated from the labeled sample only and admits the asymptotically linear representation specified in Condition $(C2')$ in the supplementary material. Under Conditions $1'-4'$ stated in the supplementary material, as $n\to\infty$, $n\leq N$, and $\sqrt{n/N}\to\rho$, we have
\begin{enumerate}
\item[(a)] $\|\widehat\beta_{\lambda}^{\mathrm{DR}}-\beta^*\|\to 0$ almost surely;
\item[(b)] $\sqrt n(\widehat\beta_{\lambda}^{\mathrm{DR}}-\beta^*)\to N(0,\boldsymbol\Sigma_{\lambda}^{\mathrm{DR}})$ in distribution;
\item[(c)] $\|\widehat\beta_{pl}^{\mathrm{DR}}-\beta^*\|\to 0$ almost surely;
\item[(d)] $\sqrt n(\widehat\beta_{pl}^{\mathrm{DR}}-\beta^*)\to N(0,\boldsymbol\Sigma_{pl}^{\mathrm{DR}})$ in distribution.
\end{enumerate}
Here
$
\boldsymbol\Sigma_{\lambda}^{\mathrm{DR}}
=(V^{\mathrm{DR}})^{-1}
\Omega_{\lambda}^{\mathrm{DR}}
(V^{\mathrm{DR}})^{-1},
$
where
$
\Omega_{\lambda}^{\mathrm{DR}}
=\{\lambda^2+(1-\lambda)^2\rho^2\}\Delta^{\mathrm{DR}}
+(1-\lambda)^2\Delta_2^{\mathrm{DR}}
+\lambda(1-\lambda)\{\Gamma^{\mathrm{DR}}+(\Gamma^{\mathrm{DR}})^{\prime}\}
+\Omega_{\eta,\lambda}^{\mathrm{DR}}.
$
The additional nuisance-estimation contribution is
\[
\begin{aligned}
\Omega_{\eta,\lambda}^{\mathrm{DR}}
={}&B_{\eta,\lambda}\operatorname{Var}(\psi_{\eta})B_{\eta,\lambda}^{\prime}
+\lambda\operatorname{Cov}(A^{\mathrm{DR}},B_{\eta,\lambda}\psi_{\eta})
+\lambda\operatorname{Cov}(B_{\eta,\lambda}\psi_{\eta},A^{\mathrm{DR}})\\
&+(1-\lambda)\operatorname{Cov}(q^{\mathrm{DR}},B_{\eta,\lambda}\psi_{\eta})
+(1-\lambda)\operatorname{Cov}(B_{\eta,\lambda}\psi_{\eta},q^{\mathrm{DR}}).
\end{aligned}
\]
For the pool estimator,
$
\boldsymbol\Sigma_{pl}^{\mathrm{DR}}
=(V^{\mathrm{DR}})^{-1}
\Omega_{pl}^{\mathrm{DR}}
(V^{\mathrm{DR}})^{-1},
$
where with $r_{\rho}=\rho^2/(1+\rho^2)$,
$
\Omega_{pl}^{\mathrm{DR}}
=r_{\rho}\Delta^{\mathrm{DR}}
+\Delta_2^{\mathrm{DR}}
+r_{\rho}\{\Gamma^{\mathrm{DR}}+(\Gamma^{\mathrm{DR}})^{\prime}\}
+\Omega_{\eta,pl}^{\mathrm{DR}},
$
and
\[
\begin{aligned}
\Omega_{\eta,pl}^{\mathrm{DR}}
={}&B_{\eta,pl}\operatorname{Var}(\psi_{\eta})B_{\eta,pl}^{\prime}
+r_{\rho}\operatorname{Cov}(A^{\mathrm{DR}},B_{\eta,pl}\psi_{\eta})
+r_{\rho}\operatorname{Cov}(B_{\eta,pl}\psi_{\eta},A^{\mathrm{DR}})\\
&+\operatorname{Cov}(q^{\mathrm{DR}},B_{\eta,pl}\psi_{\eta})
+\operatorname{Cov}(B_{\eta,pl}\psi_{\eta},q^{\mathrm{DR}}).
\end{aligned}
\]
The quantities $V^{\mathrm{DR}}$, $\Delta^{\mathrm{DR}}$, $\Delta_2^{\mathrm{DR}}$, $\Gamma^{\mathrm{DR}}$, $A^{\mathrm{DR}}$, $q^{\mathrm{DR}}$, $B_{\eta,\lambda}$, and $B_{\eta,pl}$ are defined in the supplementary matrial. A data-adaptive choice of $\lambda$ may be obtained by minimizing a plug-in or perturbation-resampling estimate of a scalar summary, such as $\operatorname{tr}(\boldsymbol\Sigma_{\lambda}^{\mathrm{DR}})$, over $[0,1]$.
In particular, $\Gamma^{\mathrm{DR}}$ is the cross covariance between the oracle DR rank-expansion term and the product-kernel imputation correction term. If the cross covariance, the imputation-correction variance, and the first-order nuisance-estimation contribution are all negligible, e.g. $\Gamma^{\mathrm{DR}}=0$, $\Delta_2^{\mathrm{DR}}=0$, and $\Omega_{\eta,\lambda}^{\mathrm{DR}}=0$, then the simple rule $\lambda=\rho^2/(1+\rho^2)=n/(n+N)$ is recovered.
\end{theorem}
Estimation of the covariance matrices in Theorem~\ref{theorem4} and proofs of the theoretical results are relegated to Section S1, and S2 of the supplementary material, respectively.

\section{Numerical studies}\label{section6}
In order to evaluate the finite sample performance of the proposed estimators with different models of outcomes and decision rules, and show that it can handle both low and high dimensional covariates, we consider a class of monotonic index models with four different cases, that is,
$$Y=\mu(\boldsymbol{X})+A D(\boldsymbol{X})+\varepsilon,$$
$(\mathrm{I})$ $\mu(\boldsymbol{X})=1+\gamma_{1}^{\prime} \boldsymbol{X}, D(\boldsymbol{X})=2 \beta_{0}^{\prime} \boldsymbol{X}$, $p$=4;\\
$(\mathrm{II})$ $\mu(\boldsymbol{X})=1+\sin \left(\gamma_{1}^{\prime} \boldsymbol{X}\right)+0.5\left(\gamma_{2}^{\prime} \boldsymbol{X}\right)^{2}, D(\boldsymbol{X})=\left(\beta_{0}^{\prime} \boldsymbol{X}\right)^{3}$, $p$=4;\\
$(\mathrm{III})$ $\mu(\boldsymbol{X})=1+\left(X_{1} X_{2}+0.5 X_{3}^{2}\right), D(\boldsymbol{X})=\left(\beta_{0}^{\prime} \boldsymbol{X}\right)^{3}$,$p$=4;\\
$(\mathrm{IV})$ $\mu(\boldsymbol{X})=1+\gamma_{1}^{\prime} \boldsymbol{X}, D(\boldsymbol{X})=2 \beta_{0}^{\prime} \boldsymbol{X}$, $p$=8;\\
where $A \sim \operatorname{Bernoulli}(0.5),$, $\boldsymbol{X} \sim N_{p}\left(0, I_{p}\right), \varepsilon \sim N\left(0,0.5^{2}\right)$, where $N_{p}(\mu, \Sigma=I_{p})$ stands for the $p$ -dimensional multivariate normal distribution with mean $\mu$, covariance matrix $\Sigma$ and $I_{p}$ denotes the $p \times p$ identity matrix. 

In cases (I)-(III), we set $p=4$ to include low- to moderate-dimensional covariates,  with $\beta_{0}=(0.5,0.5$, $-0.5,0.5)^{\prime}, \gamma_{1}=(1,-1,1,1)^{\prime}$, and $\gamma_{2}=(1,0,-1,0)^{\prime}$, respectively; in case (IV), we set $p=8$ to include somewhat high dimensional covariates, with $\beta_{0}=(0.5,0.5,-0.5,0.5,0.5,0.5,$
$-0.5,0.5)^{\prime}$, $\gamma_{1}=(1,-1,1,1,1,-1,1,1)^{\prime}$, and $\gamma_{2}=(1,0,-1,0,1,0,-1,0)^{\prime}$, respectively.

Under cases (I)-(III), the optimal treatment regime that maximizes the value function \eqref{m-eq:2.1}, denoted by $d^{\text {opt }}(\bX)=I\left(\beta_{0}^{\prime} \bX \geq c_{0}\right)$ with $c_{0}=0$ and  $\beta_0=(0.5,0.5,-0.5,0.5)^{\prime}$ after imposing the constraint $\left\|\beta_{0}\right\|=1$. Under cases (IV), the optimal treatment regime that maximizes the value function \eqref{m-eq:2.1}, denoted by $d^{\text {opt }}(\bX)=I\left(\beta_{0}^{\prime} \bX \geq c_{0}\right)$ with $c_{0}=0$ and  $\beta_0=(0.5,0.5,-0.5,0.5,0.5,0.5,$
$-0.5,0.5)^{\prime}$ after imposing the constraint $\left\|\beta_{0}\right\|=1$.
The CAL method in \cite{fan2017concordance} is denoted by FS (fully supervised), and its doubly robust extension by FS-DR.
The proposed semi-supervised method based on maximizing \eqref{m-eq:2.6} or \eqref{m-eq:2.8} is denoted by SS or SS-DR, respectively, while the proposed semi-supervised doubly robust method based on maximizing \eqref{m-eq:2.7} or \eqref{m-eq:2.9} is denoted by PL or PL-DR, respectively.

In the simulation study, we compare the proposed SS estimators ($\hat{\beta}_{\lambda}$ and $\hat{\beta}^{DR}_{\lambda}$) and the proposed PL estimators ($\hat{\beta}_{pl}$ and $\hat{\beta}^{DR}_{pl}$), with the benchmark FS estimator by \cite{fan2017concordance} ($\hat{\beta}$) that only utilizes the labeled data, to assess the relative accuracy and efficiency of our proposed estimators. For the propensity score model, we consider randomized trials with $\pi(\boldsymbol{X})=0.5.$ For each case, we carry out 200 simulation runs, and the optimization procedure is done by the optim function in $\mathbf{R}$ with the default method "Nelder-Mead" for searching the maximizer.

To evaluate the performance of different estimators for $\beta$, we report the mean bias (Bias) and standard deviation (SD) of the estimators, the mean of the estimated standard errors (SE), the empirical coverage probability (CP) of $95 \%$ percentile confidence intervals, and the efficiency(Effi) of mean square error (MSE) compared with the benchmark FS estimator, where  
$$\text{Effi}=\frac{\text { MSE of benchmark estimator }- \text { MSE of proposed estimator}}{\text { MSE of benchmark estimator }},$$ 
and the SE and CP are estimated by implementing the perturbation resampling method described in Section 5, setting the perturbation variable $G \sim \operatorname{Beta}(\sqrt{2}-1,1)$ and generating $B=200$ bootstrap samples for estimation. 
Similar results are obtained when the Exponential(1) distribution is used for perturbation.
The mean and standard deviation for $\hat{c}$ are also reported. We utilize the Gaussian kernel and set the bandwidth to be $h_n=0.5n^{-1 / 3}$ for all the cases. To assess the accuracy of the estimated optimal treatment regime $\hat{d}^{\mathrm{opt}}(\mathbf{X})=I\left(\hat{\beta}^{\prime} \mathbf{X} \geqslant \hat{c}\right)$, we report the mean and standard deviation of the percentages of making correct decisions, PCD, defined as $1-n^{-1} \sum_{i=1}^{n}\left|I\left(\hat{\beta}^{\prime} \mathbf{X}_{i} \geqslant \hat{c}\right)-I\left(\beta_{0}^{\prime} \mathbf{X}_{i} \geqslant c_{0}\right)\right| .$

The simulation results for estimators obtained under known propensity score are summarized in Table \ref{Table1} and Tables S1-S3 in the supplementary material. And the simulation results for doubly robust estimators obtained with unknown propensity scores are summarized in Tables S4-S7 in the supplementary material. In each case, the mean biases of the proposed estimators are close to zero with small SDs and corresponding CPs close to $95\%$. Our variance estimation method also performs well since the SDs are close to the SEs and the CPs are always around the predetermined significance level $95\%$. 

Especially, the efficiency improvements for all settings of SS/PL compared to FS are mostly larger than $20\%$, exhibiting the outperformance of the proposed SS/PL methods. In general, for a given size of labeled data $n=100$($n=200$), when the size of outcome-unlabeled data $N$ increases from 200 to 500 (from 400 to 1000), both the Bias and SD of the proposed SS and PL estimators reduce, with the simultaneous growth of efficiency, indicating the improvements in accuracy and efficiency for SS and PL methods with larger size of outcome-unlabeled data, respectively.  It is noticeable that under most cases, compared with FS estimators, the reduction in the bias and SD tends to be more significant, resulting in higher efficiency for PL relative to SS method. 

Further, the estimators of $c$ are closer to the true value 0 under the SS/PL methods than the FS methods with smaller SDs, showing their higher accuracy for estimation compared to the SS method. Last, the PCDs of the optimal treatment regimes obtained by the SS/PL estimators range from 0.80 to 0.94 among different cases and sample sizes and always take larger values than those obtained by the FS method, which demonstrates their advantages in decision-making for optimal treatment regimes.

\begin{table}[!htbp]
  \caption{Results under setting 1 with known propensity score, for the mean bias (Bias), standard deviation (SD) of the estimators, the mean of the estimated standard errors (SE), the empirical coverage probability (CP) of $95 \%$ confidence intervals for the estimators, and the relative efficiency(Effi) of proposed (SS/PL) estimator compared with the benchmark (FS) estimator.}
 \label{Table1}
\renewcommand{\arraystretch}{0.5}
\begin{tabular}{@{\extracolsep{0.5pt}} cccccccccc}
\hline \hline
Method&$n$&$N$&Statistics&$\hat{\beta}_1$&$\hat{\beta}_2$&$\hat{\beta}_3$&$\hat{\beta}_4$&$\hat{c}$&$PCD$ \\
\hline
FS&200&&Bias&-0.003 &0.000&  0.003 &-0.004& 0.048& 0.889\\
&&&SD&0.051 &0.052 &0.052& 0.055& 0.370& 0.080\\
&&&SE&0.056 &0.055&0.048& 0.056&-&-\\
&&&CP($\%$)&96.5 &95.0& 95.0& 95.0&-&-\\
\hline
\hline
SS&200&200&Bias&  -0.005& 0.002&  0.001& -0.003& 0.013 &0.907\\
&&&SD&0.042& 0.043 &0.045& 0.044& 0.279& 0.060\\
&&&SE&0.047 &0.047&0.044& 0.047&-&-\\
&&&CP($\%$)&95.0 &96.5& 93.5& 95.0&-&-\\
&&&Effi&0.314& 0.314& 0.230& 0.351&-&-\\
\hline
PL&200&200&Bias&-0.006 & 0.001&  -0.003& -0.002&0.013&0.911\\
&&&SD&0.031 & 0.034&  0.033 & 0.031&0.279&0.064\\
&&&SE&0.032 & 0.031&  0.031 & 0.032&-&-\\
&&&CP($\%$)&95.5 &92.5& 95.0& 95.0&-&-\\
&&&Effi&0.622 & 0.582&  0.582 & 0.664&-&-\\
\hline
\hline
SS&200&400&Bias&  -0.006& 0.001&  0.000& -0.001& 0.012 &0.920\\
&&&SD&0.038& 0.041 &0.042& 0.040& 0.241& 0.051\\
&&&SE&0.043 &0.043&0.041& 0.043&-&-\\
&&&CP($\%$)&96.0 &94.5& 95.0& 97.5&-&-\\
&&&Effi&0.447& 0.397& 0.341& 0.471&-&-\\
\hline
PL&200&400&Bias&-0.006 & 0.001&  -0.003& -0.002&0.017&0.926\\
&&&SD&0.031 & 0.034&  0.033 & 0.031&0.221&0.049\\
&&&SE&0.030 & 0.030&  0.030 & 0.030&-&-\\
&&&CP($\%$)&94.0 &91.5& 93.0& 93.0&-&-\\
&&&Effi&0.621&  0.581 & 0.582&  0.665&-&-\\
\hline
\hline
SS&200&1000&Bias&-0.005&0.000&-0.002&-0.003&0.010&0.935\\
&&&SD&0.033&0.037&0.038&0.035&0.199&0.045\\
&&&SE&0.037&0.037&0.036&0.037&-&-\\
&&&CP($\%$)&96.5&94.5&94.0&96.5&-&-\\
&&&Effi&0.576&0.514&0.449&0.600&-&-\\
\hline
PL&200&1000&Bias&-0.006 & 0.001 & -0.003 & -0.002&0.011&0.934\\
&&&SD&0.031 & 0.034&  0.033&  0.031&0.204&0.047\\
&&&SE&0.030 & 0.030&  0.030&  0.030&-&-\\
&&&CP($\%$)&94.0&91.5&93.0&93.0&-&-\\
&&&Effi&0.622&  0.581 & 0.582&  0.665&-&-\\
\hline \hline
\end{tabular}
\\
{\footnotesize $\dagger$ The true optimal regime is $d^{\mathrm{opt}}(\mathbf{x})=I\left(\beta_0^{\prime} \mathbf{x} \geqslant c_0\right)$ with $\beta_0=(0.5,0.5,-0.5,0.5)^{\prime}$ and $c_0=0$.
Here FS corresponds to the benchmark estimator by \cite{fan2017concordance} ($\hat{\beta}$) , SS corresponds to the 
proposed estimators ($\hat{\beta}_{\lambda}$) and PL corresponds to the proposed estimators ($\hat{\beta}_{pl}$).}
\end{table}

\section{Application with MIMIC-III dataset} \label{section7}
We illustrate the proposed method using the MIMIC-III database, which contains deidentified clinical records for patients admitted to critical care units at the Beth Israel Deaconess Medical Center between 2001 and 2012 \citep{johnson2016mimic, johnson2019mimic}, and is available on {\color{red}\url{https://physionet.org}.}

The MIMIC-III dataset consists of observations obtained from patients admitted to the ICU with sepsis, including routinely recorded pretreatment characteristics and treatment information. We incorporate seven baseline variables: patient age (years), weight at admission (kg), body temperature at admission (Celsius), glucose level (mg/dL), blood urea nitrogen (BUN) amount (mg/dL), creatinine amount (mg/dL), and white blood cell (WBC) count (K/uL). 
The treatment indicator is defined by whether the patient received vasopressor therapy, with $A=1$ for vasopressor use and $A=0$ for alternative medical management, such as intravenous fluid resuscitation.
The clinical outcome of interest is cumulative fluid balance (mL). A large positive fluid balance indicates fluid overload, which has been associated with adverse outcomes in critically ill patients, whereas an excessively negative fluid balance reflects insufficient circulating volume and may also be harmful \citep{claure2016fluid}. To favor fluid equilibrium, we define the response as
\[
Y=-\lvert \text{cumulative balance}\rvert,
\]
so that larger values of $Y$ correspond to more desirable outcomes.
After excluding outliers, the MIMIC-III cohort consists of 4649 patients, of whom 516 received vasopressor therapy, while the remaining patients were assigned to alternative medical management. 
For this analysis, the propensity score is estimated by logistic regression, and the conditional mean outcome model is estimated by linear regression.
Let $(\widehat\beta^{\mathrm{DR}}_{\mathrm{oracle}},\widehat c^{\mathrm{DR}}_{\mathrm{oracle}})$ denote the optimal linear treatment rule obtained from the full cohort using the CAL doubly robust procedure of \cite{fan2017concordance}; the corresponding estimated value $\hat{V}_a(\hat{\beta}_{oracle}^{\mathrm{DR}}, \hat{\alpha},\hat{c}_{oracle}^{\mathrm{DR}})$ serves as an oracle benchmark.

To mimic a semi-supervised setting, we repeatedly and randomly select 1000 patients as the labeled sample and treat the remaining patients as outcome-unlabeled, retaining their covariates and treatment assignments but masking their outcomes. For each split, we compute the optimal linear individualized treatment rule using (i) the fully supervised CAL double robust estimator based only on the labeled sample (denoted as FS-DR) and (ii) the proposed semi-supervised double robust 
(SS-DR) estimator based on both the labeled and outcome-unlabeled samples. To assess how well an estimated rule recovers the oracle rule, we also report the percentage of correct decisions,
\[
\mathrm{PCD}
=1-\frac{1}{N}\sum_{i=1}^N
\left|d(\bX_i;\widehat\beta)-d\bigl(\bX_i;\widehat\beta^{\mathrm{DR}}_{\mathrm{oracle}}\bigr)\right|.
\]
This process is repeated 50 times to ensure robustness, and Table \ref{MIMIC} summarizes the means and standard deviations of the value estimators, and the percentages of correct decisions across these 50 replications.

The results show that the ITRs estimated using the proposed SS-DR method outperform those obtained using the FS method. Specifically, the estimated values are closer to the oracle benchmark derived from the entire cohort and exhibit substantially smaller variability.
In addition, the percentage of correct decisions is higher for the SS-DR estimator.
These findings indicate that, in this electronic health record application, incorporating observed information from outcome-unlabeled data improves both the stability of estimation and the recovery of the treatment rule that would be obtained from the full dataset.
\begin{table}
\caption{Mean and standard deviations (in parentheses) of the value estimators and percentages of correct decisions}
 \label{MIMIC} 
 \begin{tabular}{cccc}
\hline\hline
\multicolumn{1}{c}{}&\multicolumn{1}{c}{Oracle}&\multicolumn{1}{c}{FS-DR}&\multicolumn{1}{c}{SS-DR}\tabularnewline
Value&-7379.45&-7420.38(278.63)&-7380.80(20.92)\\
Percentage of Correct Decisions&&99.77&99.92\\
\hline\hline
\end{tabular}
\end{table}

\section{Discussion}\label{section9}
This paper considers the SS learning for deriving optimal treatment regimes under the SS framework in many real datasets.
To address the challenge of missing outcomes and effectively leverage outcome-unlabeled data, we employ projection dimension reduction techniques in combination with kernel smoothing, allowing for flexible and accurate non-parametric imputation even for high-dimensional covariates.
The integration of information from both labeled data and outcome-unlabeled data facilitates the construction of novel SS concordance function estimators, which will be instrumental in deriving optimal treatment regimes. 

Several potential avenues for future work could enhance the scope and applicability of our findings.
In addition to the concordance function, we can explore the optimization of other types of objective functions, including value functions, and investigate non-linear treatment decisions in further studies.
In our current study, we focused on the comparison of two treatment options. Nevertheless, it is worth noting that real-world applications often involve decision-making scenarios with more than two treatment alternatives. Consequently, it is crucial to develop a more generalized method that accommodates multiple treatment options or even continuous decision variables.  Another area that holds significant promise for future investigation is the extension of our proposed SS methods to dynamic treatment regimes. 
Current methods primarily focus on a single decision time point, but there is potential for extension to incorporate multiple decision time points, with subjects receiving treatments sequentially, and intermediate outcomes influencing subsequent treatment choices. This can be achieved by utilizing the class of monotonic index models and employing the single index kernel smoothing techniques. 

A key assumption is made in this paper that the labeled data and outcome-unlabeled data have the same joint distribution of $(\bX,A)$ to validate the proposed methods. Otherwise, the optimal treatment regime that maximizes the concordance function for the labeled data and outcome-unlabeled data may be different. If the covariate or treatment-assignment distributions are different, some calibration is needed. Recently, a lot of methods (e.g., using density ratio as weights or using the inverse probability weighting technique) have been proposed to handle covariate shifts when combining different data sources. 
 \cite{cai2021coda} proposed a calibrated value estimator by rebalancing the value estimators of common intermediate outcomes in two samples based on their posterior sampling probability when the distributions of baseline covariates differ in different samples. 
 \cite{li2021targeting} proposed a calibrated augmented inverse probability weighted estimator of the value function for the target population
and proposed a methodology framework that integrates heterogeneous data from diverse populations through a two-way data integration strategy.
\cite{mo2021learning} proposed a distributionally robust ITR (DR-ITR) framework that maximizes the worst-case value function across the values under a set of underlying distributions that are ``close” to the training distribution, where there are some unknown covariate changes between the training and testing distributions.
\cite{chu2022targeted} proposed a calibrated augmented inverse probability weighted estimator of the value function for the target population with only summary information, whereas the source and target populations may be heterogeneous, and individual data is only available from the source population.
By learning from the insights gained through calibration methods and integrating them with the proposed SS approaches, a potential solution can be formulated to tackle the challenge of heterogeneity.

In practical scenarios, there is often auxiliary information available that can be valuable in enhancing the derivation of optimal treatment regimes.
\cite{zhang2022prior} introduced the prior adaptive SS (PASS) estimator, which incorporates prior knowledge by adaptively shrinking the estimator towards a direction derived from the prior.
Utilizing the additional auxiliary variable in the training dataset, \cite{xie2021nonparametric} proposed a two-step nonparametric estimation of conditional expectation, with sufficient dimension reduction for kernel estimation in both steps.
By considering and integrating auxiliary information into the SS analysis, researchers can extract additional insights and improve the overall quality of treatment recommendations in practical settings.

We believe efforts in these directions will expand the applicability of our research and enable more robust and insightful decision support in a wider range of practical scenarios. We leave them for future research.

\section{Disclosure statement}\label{disclosure-statement}

The authors have the following conflicts of interest to declare (or
replace with a statement that no conflicts of interest exist).

\section{Data Availability Statement}\label{data-availability-statement}

ACTG175 dataset is available in the R package ``BART''.
MIMIC-III dataset is available at the following URL: https://physionet.org.

\phantomsection\label{supplementary-material}
\bigskip

\begin{center}

{\large\bf SUPPLEMENTARY MATERIAL}

\end{center}

\begin{description}
\item[CAL-SS-supplementary.pdf]
Title: Supplementary Material for “Semi-supervised Concordance Learning for Optimal Individual Treatment Regimes”

\end{description}

  \bibliography{bibliography.bib}

\end{document}